\documentclass[aps,superscriptaddress,amsmath,amssymb,floatfix,twocolumn,prl]{revtex4-1}
\usepackage[]{standalone}
\usepackage{times}
\usepackage{graphicx}
\usepackage{color}
\usepackage{bm}
\usepackage{braket}
\usepackage{mathtools}
\usepackage{mathcomp}
\usepackage[caption=false,justification=raggedright,position=top,singlelinecheck=off]{subfig}
\usepackage{soul}

\newcommand{\bk}{\bm{k}}
\newcommand{\bq}{\bm{q}}
\newcommand{\bQ}{\bm{Q}}
\newcommand{\bs}{\bm{s}}
\newcommand{\bS}{\bm{S}}

\newcommand{\loc}{\text{loc}}
\newcommand{\pdag}{\phantom{\dagger}}

\begin{document}
\title{Quantum criticality and global phase diagram of an Ising-anisotropic Kondo lattice}

\author{Emilian M.\ Nica}
\email[Corresponding author: ]{en5@rice.edu}
\affiliation{Department of Physics and Astronomy  \& Rice Center for Quantum Materials, 
Rice University, Houston, Texas 77005}

\author{Kevin Ingersent}
\affiliation{Department of Physics, University of Florida, P.O. Box 118440, Gainesville,
Florida 32611}

\author{Qimiao Si}
\affiliation{Department of Physics and Astronomy  \& Rice Center for Quantum Materials, 
Rice University, Houston, Texas 77005}

\date{\today}

\begin{abstract}
Recent studies of heavy-fermion systems with tunable quantum fluctuations have focused on
a variety of zero-temperature phase transitions that involve not only the onset of magnetic
order but also the destruction of Kondo entanglement. Motivated by these developments, we
investigate
the effect of enhanced quantum fluctuations induced by a transverse magnetic field
in an
Ising-anisotropic Kondo lattice model, solved within
an extended dynamical mean field theory
using the
numerical renormalization
group. A
line of locally critical
points describes
a direct transition from a
Kondo-screened paramagnetic heavy-fermion state to a Kondo-destroyed antiferromagnetic
phase. Along the line, the extracted critical exponents remain unchanged.
By probing the the interplay between
quantum fluctuations of the local moments,
the
Kondo effect, and magnetic order,
this study provides significant new insight
into the global phase diagram of heavy-fermion systems.
\end{abstract}

\maketitle

Heavy-fermion materials have provided
condensed matter
physics
with some of its most challenging yet fascinating problems. In the recent past,
these materials
have also presented a canonical setting to study non-Fermi liquid  behavior in the context
of quantum criticality~\cite{NFL_Rev}. Inspired by the unusual quantum critical properties
of prototypical systems such as YbRh$_2$Si$_2$~\cite{Paschen_et_al_Nat_2004,
Gegenwart_et_al_Science_2007,Friedemann_PNAS_2010},
CeCu$_{6-x}$Au$_x$~\cite{Schroder_et_al_Nat_2000},
and CeRhIn$_5$ ~\cite{Park_Nat2006,Shishido_et_al_JPSJ_2005},
several authors~\cite{Global_phase_diagram,Coleman_Nevid2010} proposed
the global phase diagram (GPD) sketched in Fig.~\ref{Fig:Glbl_phs_dgrm}(a).
In the form relevant to
the present
work, the GPD addresses the fundamental role of quantum fluctuations and their effects on
the basic Doniach picture~\cite{Doniach_PRB_1977}:
in addition to the ratio of the bare Kondo scale $T_K^{(0)}$
to the RKKY interaction scale $I$ [the horizontal axis of
Fig.~\ref{Fig:Glbl_phs_dgrm}(a)],
there is a second axis characterized
by a quantity $G$ that
measures the quantum fluctuations of the local-moment magnetism.
The GPD naturally incorporates Kondo destruction and the associated local quantum
criticality~\cite{Si_et_al_Nature_2001, Si_et_al_PRB_2003}, which describes a direct transition
(at temperature $T=0$) between (i) a paramagnetic phase with a large Fermi surface
$\left(P_L\right)$ induced by the Kondo effect and (ii) an antiferromagnetically
ordered phase with a small Fermi surface $\left(AF_S\right)$
produced by Kondo destruction.
Such a phase diagram appears pertinent to a variety of strongly correlated systems
beyond heavy fermions, such as the organic charge-transfer salts \cite{Kanoda} and
high-$T_c$ cuprates \cite{Taillefer}.

The GPD has had considerable influence on the understanding and classification of
quantum critical heavy-fermion
metals with respect to variations of
dimensionality~\cite{Custers-2012}, geometrical frustration~\cite{Fritsch-2014,Kim-2013,
Khalyavin-2013,Mun-2013,Tokiwa-2015}, or otherwise tunable quantum
fluctuations~\cite{Jiao-2015,Tomita2015,Friedemann-2009,
Custers-2010}. However, there have been very few
concrete theoretical calculations
investigating
the nature of the phase transitions.
An important open question~\cite{Global_phase_diagram,Coleman_Nevid2010,
Nica_et_al_PRB_2013} is whether there is indeed a line of local critical
points where 
the Kondo destruction
occurs simultaneously
with the formation of long-range order, as crossed by the Type-I trajectories
shown in Fig.~\ref{Fig:Glbl_phs_dgrm}(a),
or only a single point that is multi-critical.

In this work, we present the first theoretical study of this outstanding
issue, carried out by incorporating
into an Ising-anisotropic Kondo lattice model the effect of tunable quantum
fluctuations induced by a transverse magnetic field.
Our main result, summarized in Fig.~\ref{Fig:Glbl_phs_dgrm}(b), is the existence
of a line of continuous type-I transitions.
Along this line, the critical behavior is shown to exhibit little variation for
transverse fields smaller than a bare Kondo scale, with the dynamical properties
satisfying a remarkable superscaling.

\emph{Model and solution method:}
We study a Kondo lattice model
with Ising anisotropy of the spin exchange between local moments,
and tune the quantum fluctuations of the local moments through
an external transverse field.
The Hamiltonian reads
\noindent \begin{align}
\label{Eq:KLM}
H_{\text{KLM}}= & \sum_{ij} t_{ij} \; c^{\dagger}_{i \sigma} c^{\pdag}_{j \sigma}
+ J_{K} \sum_{i} \bS_{i} \cdot \bs_{i} \notag  \\
& +\frac{1}{2} \sum_{ij} I_{ij} S_{i}^{(z)} \cdot S_{j}^{(z)} + \Delta \sum_{i}  S^{(x)}_{i}.
\end{align}
\noindent
The first term represents the kinetic energy
of a band of spin-$\frac{1}{2}$ conduction
while the second
describes an $SU(2)$-invariant Kondo coupling at each site $i$
between a local impurity spin
$\bS_{i}$ and the on-site spin density of the conduction electrons
$\bs_{i}$.
The third
term in $H_{\text{KLM}}$ is a exchange coupling between the $z$ components of
the local moments.
Although this coupling is generated by the previous terms
to second-order in the Kondo exchange and corresponds to the
Ruderman-Kittel-Kasuya-Yosida (RKKY) interaction,
we include it separately
in order to tune the magnetic fluctuations.
The last term in Eq.~\eqref{Eq:KLM} introduces the transverse field $\Delta$.
We note that, (i) the role of transverse field in the (local-moment) quantum Ising model
is well studied \cite{Rosenbaum-1996,TFIM-Chakravarti1996};
(ii) for two-dimensional magnetic fluctuations in the absence of the transverse
field, the Ising-anisotropic
Kondo lattice
has been shown to display a locally critical point~\cite{Si_et_al_Nature_2001,
Si_et_al_PRB_2003,Grempel_Si_PRL_2003,Zhu_et_al_PRL_2003,Glossop_Ingersent_PRL_2007,Zhu_et_al_PRL_2007};
(iii) an easy-axis anisotropy is observed in a number of heavy-fermion systems
~\cite{Schroder_et_al_Nat_2000,Kim-2013,Stockert-2007,Aronson-2014}.

\begin{figure*}[t!]
\subfloat[]{\includegraphics[width=0.6\columnwidth]{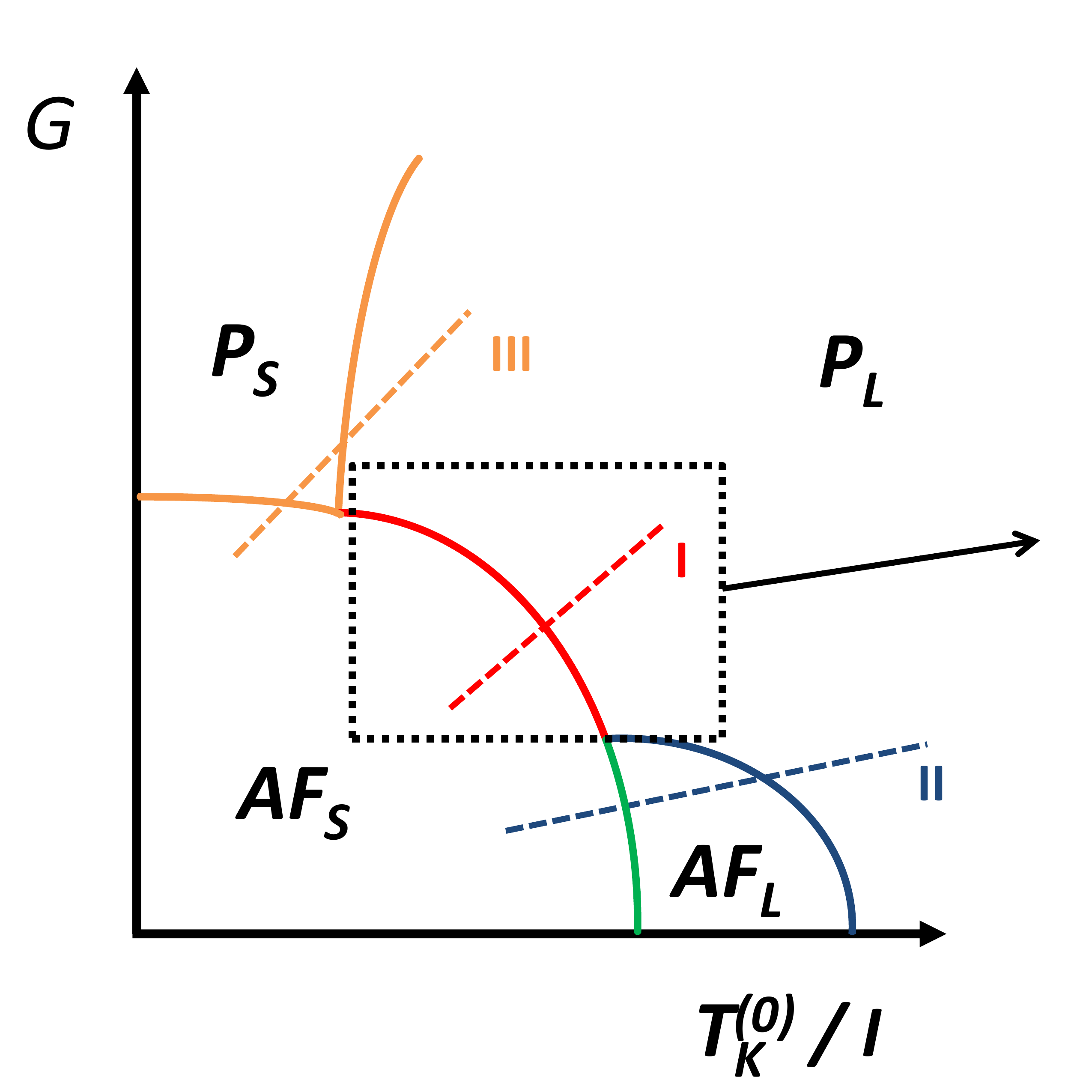}} \qquad
\subfloat[]{\includegraphics[width=0.64\columnwidth]{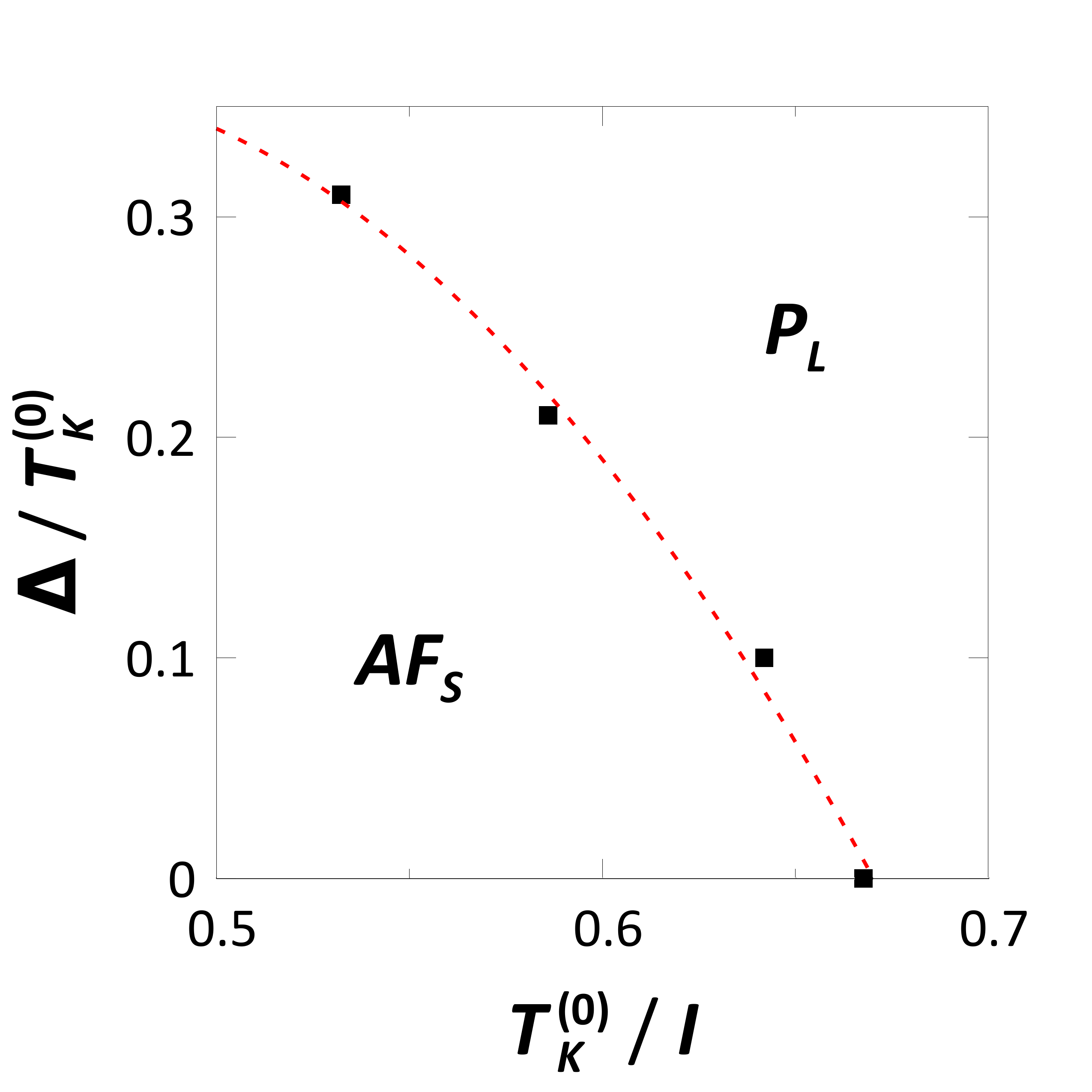}}
\caption{(a) Proposed global phase diagram~\cite{Global_phase_diagram}
motivating this work.
The horizontal axis represents the competition between
the Kondo coupling, parametrized by the bare Kondo temperature $T_{K}^{(0)}$
and the RKKY coupling $I$. The vertical axis plots a quantity $G$ measuring
the degree of quantum fluctuations of the local moments.
$P$ and $AF$ label paramagnetic and antiferromagnetic phases while the
indices $L$ and $S$ denote large and small Fermi surfaces, respectively.
The
rectangle indicates the region of focus of this study.
(b) Phase diagram of the Ising-anisotropic Kondo lattice
as found in this work. It can be regarded as a concrete realization of the
rectangular region of (a) with $G=\Delta/T_{K}^{(0)}$, where $\Delta$ is a transverse
magnetic field that tunes the quantum fluctuations of the local moments.
}
\label{Fig:Glbl_phs_dgrm}
\end{figure*}

We study the Hamiltonian
\eqref{Eq:KLM}
using the extended dynamical mean-field theory (EDMFT) approach~\cite{Smith_Si_PRB_2000}.
The conduction-electron Green's function and the lattice susceptibilities take the forms (for real frequencies)
\begin{align}
G_{\sigma}(\bk, \omega)
&= [ \omega- \epsilon_{\bk}-\Sigma_{\sigma}({\omega}) ]^{-1},
\\[1ex]
\label{Eq:Lttc_sscptblt}
\chi(\bq, \omega)
&= [ I_{\bq}+M(\omega)]^{-1}.
\end{align}
Here, $\epsilon_{\bk}$ and $I_{\bq}$ are
Fourier transforms of the hopping matrix $t_{ij}$ and
the RKKY coupling $I_{ij}$ in Eq.~\eqref{Eq:KLM},
while $\Sigma_{\sigma}(\omega)$ and $M(\omega)$ are the single-particle and spin
self-energies.
The local susceptibility is given by
\begin{equation}
\label{Eq:Slf_cnsstnc_1}
\chi_{\loc}(\omega)=\sum_{\bq}\chi(\bq, \omega).
\end{equation}
Within the EDMFT, it can be obtained from a \emph{single-impurity} Bose-Fermi Kondo
model (BFKM)~\cite{Si_et_al_PRB_2003,Zhu_Si_PRB_2002,Zarand_2012,Glossop_Ingersent_PRL_2005}
having the Hamiltonian
\begin{align}
\label{Eq:H_BFKM}
H_{\text{BFKM}}= & \sum_{p} \epsilon_{p} \, c^{\dagger}_{p \sigma} c^{\pdag}_{p \sigma}
+ J_{K} \bS \cdot \bs_{0} \notag  \\
& +\sum_{l} w_{l} \phi^{\dagger}_{l} \phi^{\pdag}_{l} +  S^{(z)} \sum_{l} g_{l}
\left(\phi^{\pdag}_{l} + \phi^{\dagger}_{-l} \right)
\notag \\
& + \Delta S^{(x)} + h_{\loc}S^{(z)},
\end{align}
where we introduced a bosonic field
$\phi_{p}$,
which couples to the local moment through
$g_{p}$. Note that the $p$ and $l$ indices should not be confused with the
physical momenta $\bk$ and $\bq$.
The local coupling to the bosonic bath can be parametrized~\cite{Grempel_Si_PRL_2003}
by the Weiss field $\chi_{0}^{-1}(\omega)$ satisfying
\noindent \begin{equation}
\label{Eq:Wss_2}
\text{Im}\chi_{0}^{-1}(\omega)
= \text{sgn}(\omega) \pi \sum_{l} g^{2}_{l}  \delta(\omega-\omega_{l})
= \text{sgn}(\omega) \,B(\omega),
\end{equation}
which also defines the spectral function $B(\omega)$~\cite{Glossop_Ingersent_PRL_2005}.
More details are given in the Supplemental Material~\cite{supp}.
The Weiss field satisfies the Dyson equation
\noindent \begin{equation}
\label{Eq:Wss_1}
\chi_{0}^{-1}(\omega)= M(\omega)-\chi_{\loc}^{-1}(\omega).
\end{equation}
Equation~\eqref{Eq:H_BFKM} also contains a local longitudinal field $h_{\loc}$,
which is nonzero for the magnetically ordered solution and satisfies
\begin{equation}
h_{\loc}= -\left[I -\chi_{0}^{-1}(\omega=0) \right]m_{\text{AF}}
\label{Eq:Slf_cnsstnc_2},
\end{equation}
where $m_{\text{AF}}=\braket{S^{(z)}}$ is the order parameter.
In our calculations, we take an RKKY density of states
$\rho_I(\omega)=(2I)^{-1}\Theta(I-|\omega|)$ corresponding to a generic
case of two-dimensional magnetic fluctuations. This implies
that the most negative value of the RKKY coupling is
$I_{\bq=\bQ} =-I$
occurring at some ordering wave-vector $\bQ$;
using Eqs.~\eqref{Eq:Lttc_sscptblt} and \eqref{Eq:Slf_cnsstnc_1}
we obtain the self-consistency condition~\cite{Zhu_et_al_PRL_2003}
\begin{equation}
 \label{Eq:Slf_cnsstnc_3}
\chi_{\loc}(\omega)
= \frac{1}{2I} \, \ln \left[1+2I\chi(\bQ, \omega) \right].
\end{equation}
For generic conduction-electron fillings that correspond to the case of
heavy-fermion metals, the EDMFT
uses a fixed, featureless conduction-electron density of states
$\rho(\epsilon)=(2D)^{-1} \, \Theta(D-|\epsilon|)$
to avoid double-counting the RKKY
interactions~\cite{Si_Zhu_Grempel_J_Phys_Condens_Matter_2005}. This
usage is supported by analytical studies~\cite{Zhu_Si_PRB_2002} indicating that
corrections to $\Sigma_{\sigma}(\omega)$ generate subleading terms close to the
critical point. In the case of local criticality\cite{Si_et_al_PRB_2003},
a singularity in the single-particle spectrum occurs when the local
susceptibility is divergent, which marks a critical destruction of Kondo screening.

Solution of the EDMFT equations requires
iterated solution of the impurity BFKM Hamiltonian
\eqref{Eq:H_BFKM} until the Weiss field $\chi_{0}^{-1}(\omega)$ converges
to within a predefined tolerance.
For each trial form of $\chi_{0}^{-1}(\omega)$, the impurity problem must be solved
using a Bose-Fermi variant of the numerical renormalization
group~\cite{Glossop_Ingersent_PRL_2005,Nica_et_al_PRB_2013}.
In this work each single EDMFT iteration
required several hours of running time.
While self-consistent solutions far from criticality converged within 20 iterations
or fewer, solutions closest to quantum phase transitions typically required 150-250
iterations.
Further details of the numerical procedure are given in the Supplemental Material~\cite{supp}.

\noindent \begin{figure}[t!]
\includegraphics[width=1.0\columnwidth]{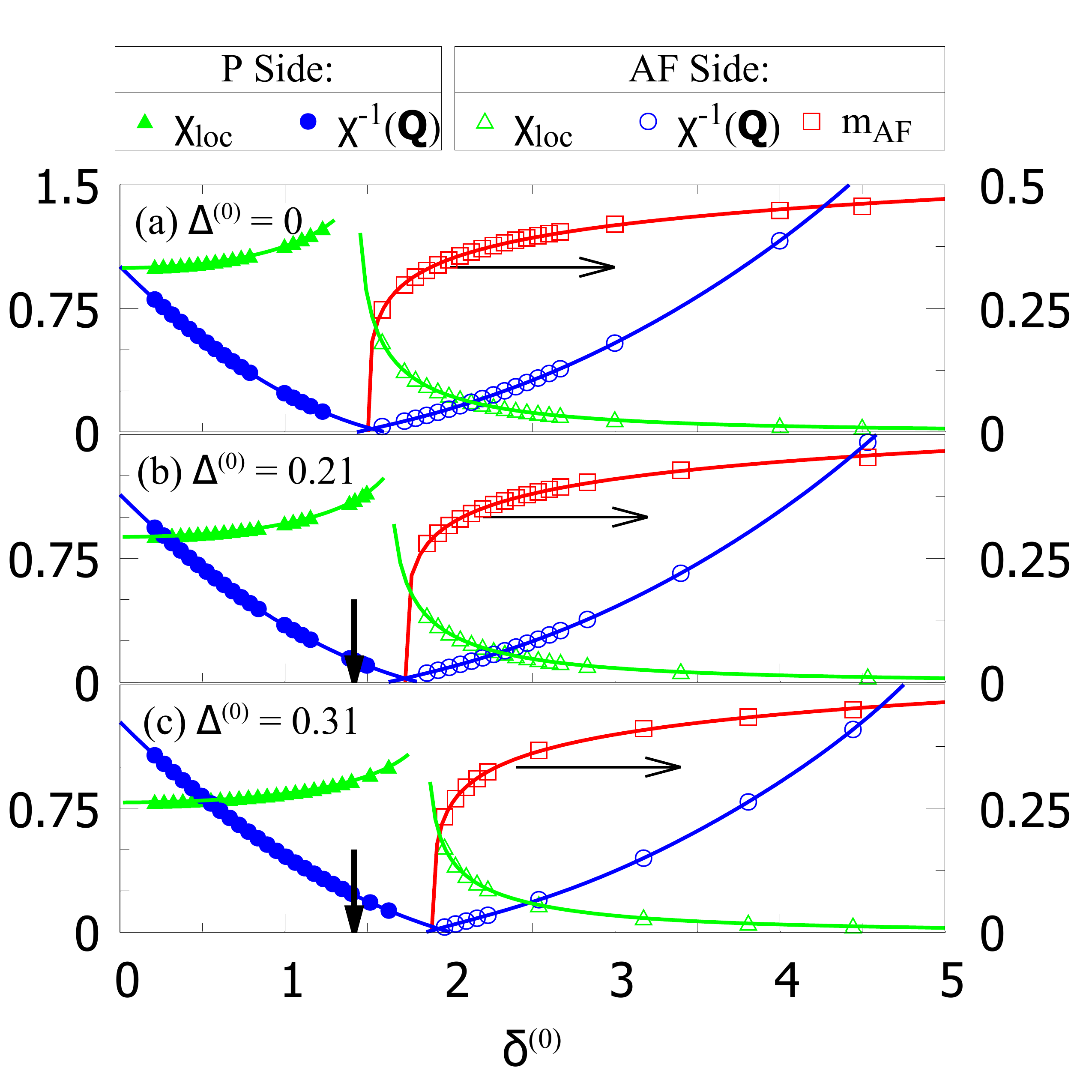}
\caption{
Quantum phase transitions of the Kondo lattice model
arising in EDMFT solutions for values of the dimensionless transverse field
(a) $\Delta^{(0)}=0$, (b) $\Delta^{(0)}=0.21$, and (c) $\Delta^{(0)}=0.31$.
As the dimensionless RKKY coupling $\delta^{(0)}$ is tuned toward
the critical point at $\delta_{c}^{(0)}$, the lattice static inverse
susceptibility $\chi^{-1}({\bf Q},\omega=0)$ vanishes and the local static
susceptibility $\chi_{\loc}(\omega=0)$ diverges
from both the paramagnetic ($P$) and the ordered ($AF$) sides.
Accompanying this, the order-parameter $m_{\text{AF}}$ vanishes as $\delta^{(0)}$
approaches $\delta_{c}^{(0)}$ from the $AF$ side.
The solid lines are
guides to the eye.
In (b) and (c), the arrows mark the
critical coupling for the zero transverse-field
case
in (a), as extrapolated from the ordered side.
 }
\label{Fig:EDMFT_multiplot_scaled_by_Del_0}
\end{figure}

\emph{Line of critical points and scaling of critical dynamics}:
For each value of the transverse field $\Delta$,
we study the quantum phase transition by tuning the
RKKY coupling $I$ with a fixed Kondo coupling $J_K$, or equivalently,
a fixed bare Kondo scale defined in terms of the local static susceptibility
as $T_{K}^{(0)}=\chi_{\loc}^{-1}(\omega=0; T=I=\Delta=0)$.
We
express the transverse field and the RKKY coupling in units of $T_{K}^{(0)}$,
\textit{i.e.}, by defining $\Delta^{(0)} \equiv \Delta /T_K^{(0)}$ and
$\delta^{(0)} \equiv I /T_K^{(0)}$.
We
analyze the representative cases $\Delta^{(0)}= 0$, $0.1$, $0.21$, and $0.31$.
Our discussion focuses on the two largest values of the transverse field,
but very similar results are
obtained for $\Delta^{(0)}=0.1$.

Figure~\ref{Fig:EDMFT_multiplot_scaled_by_Del_0} shows the evolution of the
thermodynamics in the limit $T \rightarrow 0$
as functions of the normalized RKKY coupling $\delta^{(0)}$ for $\Delta^{(0)}=0$, $0.21$,
and $0.31$.
The local and lattice susceptibilities diverge as one approaches the critical
point at $\delta^{(0)} = \delta^{(0)}_c$ from either the $P$ or the $AF$ phase. On
approach from the
ordered side, the
magnetic order-parameter $m_{\text{AF}}$ also appears to vanish continuously.
There is a narrow region of coexistence of ordered and disordered solutions.
The critical couplings $\delta_{c,P}^{(0)}$ and $\delta_{c,\text{AF}}^{(0)}$ obtained
via linear extrapolation of $\chi^{-1}(\bQ)$ from both sides differ by less than 5\% of
the mean critical coupling.
Similar to previous $\Delta=0$ numerical renormalization-group
results~\cite{Glossop_Ingersent_PRL_2007, Zhu_et_al_PRL_2007},
including their Landau analysis~\cite{Zhu_et_al_PRL_2007}, and other
analytical and numerical studies~\cite{Si_et_al_PRB_2003, Grempel_Si_PRL_2003,Zhu_et_al_PRL_2003},
these thermodynamic
properties for $\Delta \neq 0$ as well as
the dynamic scaling behavior (see below)
suggest the second-order nature of the $T=0$ transition.
These results establish the existence of $P_L$-$AF_S$ transitions in the presence of a
transverse field with a phase boundary shown in Fig.~\ref{Fig:Glbl_phs_dgrm}(b).

To demonstrate that the transitions for nonzero $\Delta^{(0)}$ are most
likely of the locally critical type, and to provide further evidence against first-order
transitions, we plot in Fig.~\ref{Fig:Lcl_crtcl}(a) the inverse local susceptibility
versus the inverse of the lattice susceptibility at $\omega=T=0$
for $\Delta^{(0)}=0.21$ on the $P$ side.
The calculated values (squares) match the dimensionless function
$y=2/\text{ln}(1+2/x)$ (line, from Eq.~(\ref{Eq:Slf_cnsstnc_3}));
this indicates that the two susceptibilities diverge simultaneously at the
critical point, as expected
for local criticality with two-dimensional magnetic fluctations.
Similar results were obtained for $\Delta^{(0)}=0, 0.1$, and 0.31.

We also analyze the dynamical scaling of the $T=0$ local susceptibility,
which on approach to the critical point is expected to
take a form similar to its $\Delta^{(0)}=0$
counterpart~\cite{Si_et_al_PRB_2003,Glossop_Ingersent_PRL_2007}:
\begin{align}
\label{Eq:Ch_ln_sclng}
T_{K}^{(i)} \: \text{Re} \left[ \chi_{\loc} (\omega) \right]
 \sim \frac{\alpha}{2\delta^{(i)}_c} \: \ln \left| \frac{T_K^{(i)}}{\omega} \right|.
\end{align}
Here, we have defined the bare Kondo scale
$T_K^{(i)}=\chi^{-1}_{\loc}(\omega=0; T=I=0, \Delta=\Delta^{(0,i)}\,T_K^{(0)})$
in the presence of a scaled transverse field $\Delta^{(0,i)}=0$, $0.21$, and
$0.31$ for $i=0$, $1$, and $2$, respectively. In Eq.~\eqref{Eq:Ch_ln_sclng},
$\alpha( \delta_c^{(i)}, \Delta^{(i)})$ characterizes the frequency dependence
$\chi({\bf Q}, \omega; T=0) \propto \omega^{-\alpha}$ of the lattice susceptibility
at the critical value $\delta_c^{(i)}$ of $\delta^{(i)}=I/T_K^{(i)}$.

\begin{figure}[t!]
\includegraphics[width=1\columnwidth]{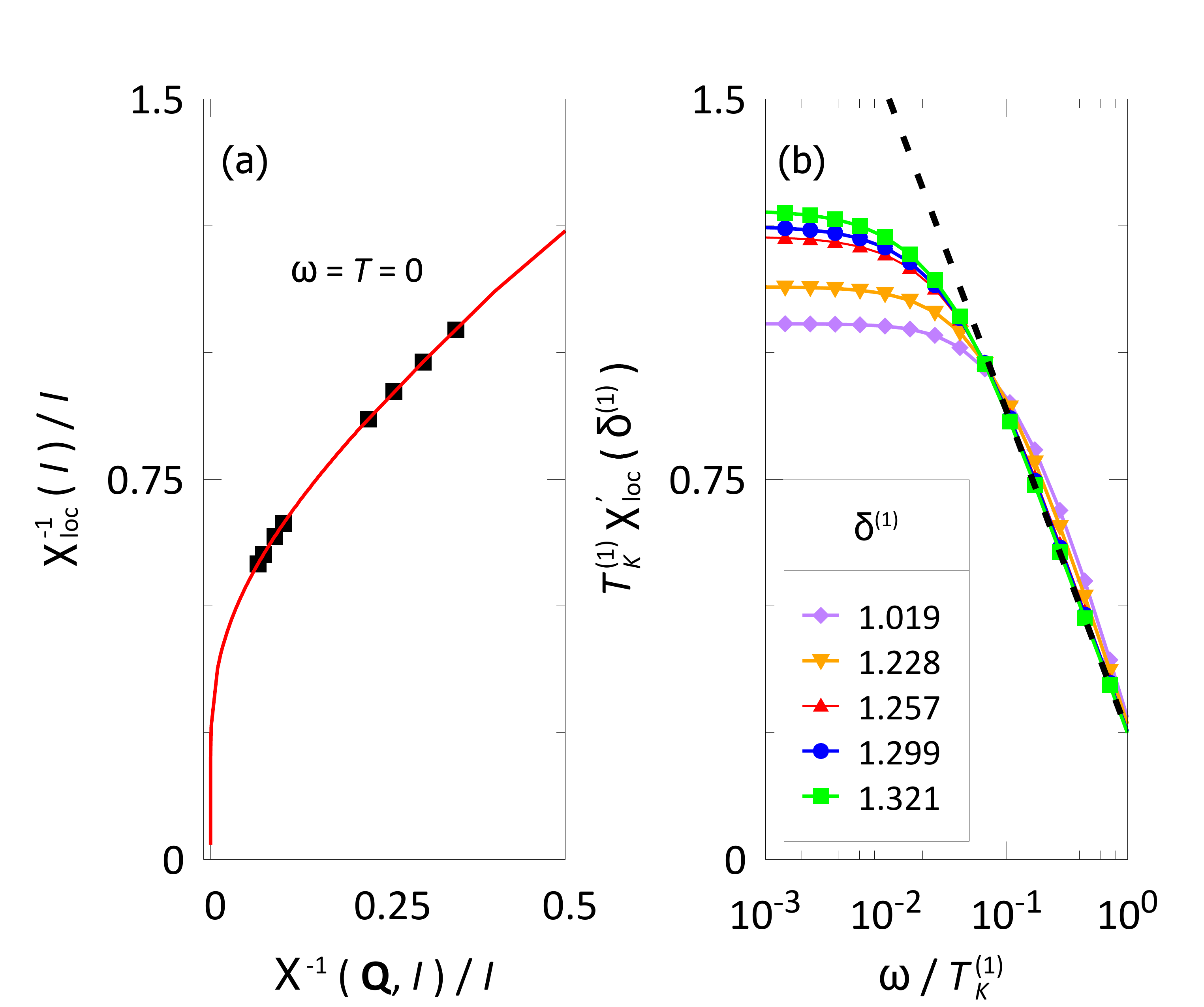}
\caption{(a) Inverse local susceptibility vs inverse lattice susceptibility at
$\omega=T=0$ for scaled transverse field $\Delta^{(0)}=0.21$ on the approach to
the transition from the $P$ side. The squares show the calculated values while the
solid line is is determined by Eq.~\eqref{Eq:Slf_cnsstnc_3} with appropriate
normalization, as discussed in the main text. The match between the expected and
calculated values indicates that the two susceptibilities diverge simultaneously
at the extrapolated critical coupling and that the transition is of
the second-order, locally critical type.
(b) Semi-log plot of the real part of the local susceptibility vs frequency at $T=0$
for increasing values of $\delta^{(1)}$ from the $P$ side. Note that all quantities
have been scaled by $T_{K}^{(1)}$.
A logarithmic form (broken line) given by Eq.~\eqref{Eq:Ch_ln_sclng}
is approached as $\delta^{(1)} \rightarrow \delta^{(1)}_{c}=1.507$.
}
\label{Fig:Lcl_crtcl}
\end{figure}

Figure \ref{Fig:Lcl_crtcl}(b) is a semi-log plot of the real part of the local
susceptibility at $T=0$ versus frequency for several values of
$\delta^{(1)} < \delta^{(1)}_{c,P}$ ($P$ side) for $\Delta^{(0)}=0.21$.
As
$\delta^{(1)}$ approaches its critical value $\delta_c^{(1)}=1.507$, the curves tend to
a logarithmic dependence over an intermediate frequency range.
This indicates that the logarithmic form typical of a locally critical solution for $\Delta^{(0)}=0$
is also present for finite transverse fields, and we conclude that we are again
seeing a locally critical transition.
The value of the extracted exponent [Eq.~\eqref{Eq:Ch_ln_sclng}] is $\alpha=0.85\pm 0.03$.
We can understand the difference between our estimate and the value $\alpha=0.78(4)$ found
for $\Delta^{(0)}=0$ \cite{Glossop_Ingersent_PRL_2007} as a consequence of different
Wilson parameters used in the numerical discretization of the band and bath continua:
$\Lambda=3$ used previously vs $\Lambda=9$ here.
Based on a superscaling discussed below [see Fig.~\ref{Fig:Sprsclng}(a)],
we identify the value of $\alpha$ for nonzero $\Delta^{(0)}$ with that for the
zero-transverse field case. A linear extrapolation in terms of $\log(\Lambda)$
(Ref.~\onlinecite{Glossop_et_al_unpub_2013} and Supplemental Material~\cite{supp})
estimates a continuum-limit value $\alpha (\Lambda \rightarrow 1)=0.73 \pm 0.01$.
This compares well with the exponent $\alpha\approx 0.75$ extracted from
neutron-scattering measurements in CeCu$_{5.9}$Au$_{0.1}$~\cite{Schroder_et_al_Nat_2000}.

We next turn to the analysis of the scaling for the lattice spin susceptibility.
A previous study of the standalone BFKM revealed a critical spectrum that varies
continuously with the applied transverse-field, strongly suggesting that a marginal
operator controls the renormalization-group flow along the critical line~\cite{Nica_et_al_PRB_2013}.
As the self-consistent BFKM solutions form a subset of possible solutions
of the standalone BFKM, a similar marginal behavior is the most likely scenario
in the EDMFT case. This has motivated us to analyze the relationship of the critical
susceptibility for the varying $\Delta$.
We find that $\chi_{\loc}(\omega)$ along the critical line collapses onto a superscaling
form, when $\chi_{\rm loc}$ and $\omega$ are both normalized in terms of the
field-dependent bare Kondo scale $T^{(i)}_K$ defined above.
The scaling collapse is demonstrated in Fig.~\ref{Fig:Sprsclng}(a), where
for each value of $\Delta^{(0)}$, $\delta^{(i)}$ is chosen closest to its extrapolated
critical value $\delta^{(i)}_{c,P}$. The small deviations from the scaling form
can be attributed to differences in $\delta^{(i)}-\delta^{(i)}_{c}$ between the three cases.
The observed superscaling shows that the critical behavior stays the same along the line of transitions,
which in turn is compatible with the proposed marginality.

To further illustrate this point, Fig.~\ref{Fig:Sprsclng}(b) plots the static data
from Figs.~\ref{Fig:EDMFT_multiplot_scaled_by_Del_0}(a)--(c)
as functions of $\delta^{(i)}$ rather than $\delta^{(0)}$. The solid lines
represent fits to the zero-transverse-field data.
Good data collapse is apparent as the transition is approached from either side,
implying very close values of $\delta^{(i)}_{c}$ for the cases $i=0$, $1$, and $2$.
Marked deviations are seen only on the AF side far from the critical points.

\noindent \begin{figure}[t!]
\includegraphics[width=1\columnwidth]{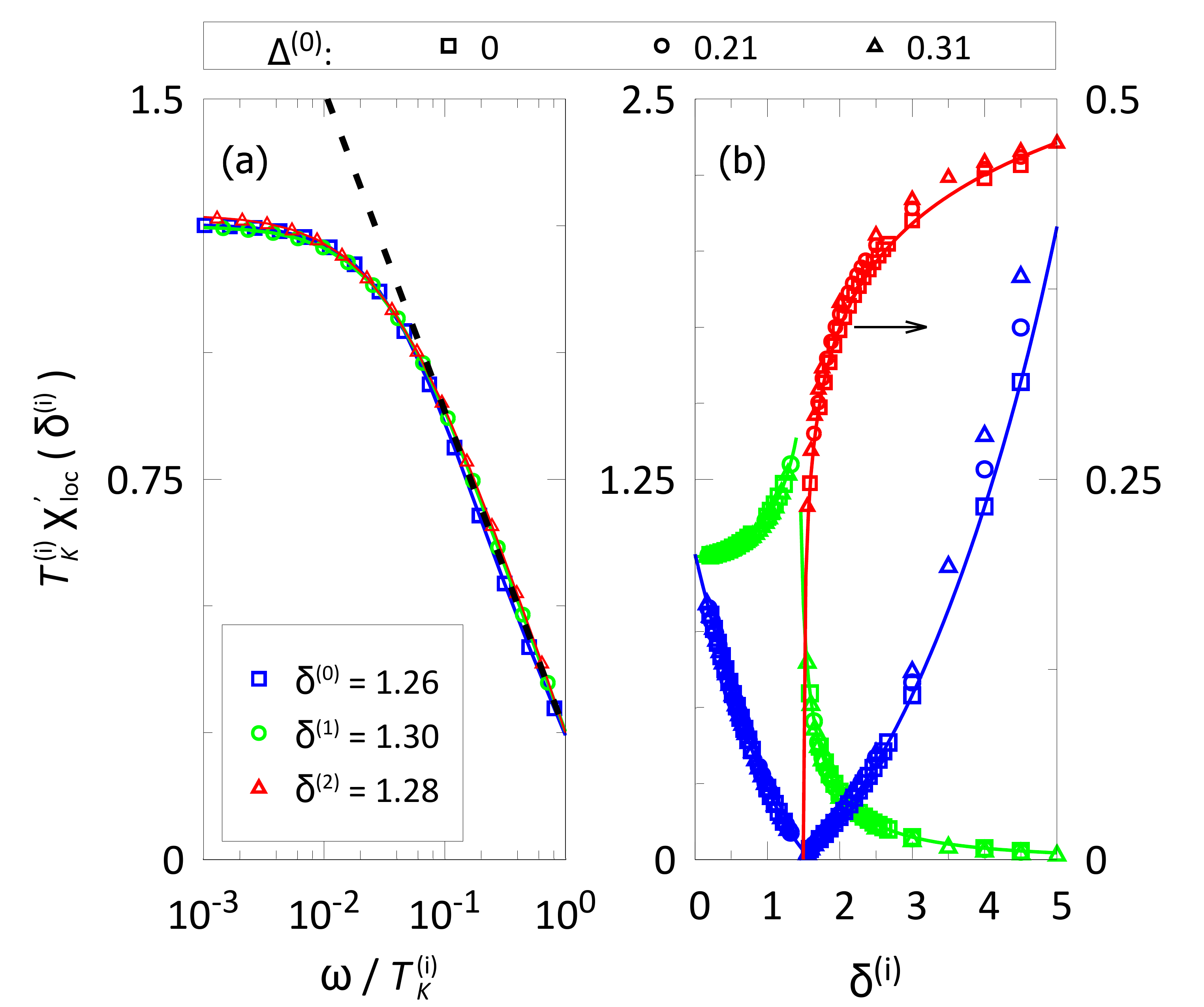}
\caption{(a) Scaling collapse on a semi-log plot of the real part of the $T=0$ local
susceptibility vs frequency for $\Delta^{(0)}=0, 0.21$, and $0.31$. In each case, the data
shown are for the $\delta^{(i)}=I/T_{K}^{(i)}$ closest to its critical value $\delta^{(i)}_{c}$.
(b) Scaling collapse for the thermodynamic data shown
in Figs.~\ref{Fig:EDMFT_multiplot_scaled_by_Del_0}(a)-(c) as functions of the appropriately
scaled tuning parameter $\delta^{(i)}$.
}
\label{Fig:Sprsclng}
\end{figure}

We close with several remarks. Firstly, our results provide a concrete realization of
a line of type-I transitions in the proposed global phase diagram in Fig.~\ref{Fig:Glbl_phs_dgrm}.
They also demonstrate a superscaling of the magnetic response that implies
that the critical exponents are unchanged along this line.

Secondly, the Kondo lattice model [Eq.~\eqref{Eq:KLM})] in the limit of $J_K \rightarrow 0$
exhibits, for sufficiently large $\Delta$, a spin-gapped paramagnetic phase
\cite{TFIM-Chakravarti1996}.
We therefore expect the emergence of a $P_S$ phase for large $\Delta$ and small $J_K$.
However, this regime is difficult to study, because the
$\Delta$-induced transition at $T=0$ in the absence of any coupling to fermions
is first-order in EDMFT (see the Supplemental Material~\cite{supp}).
Indeed, for values of $\Delta^{(0)}$ approaching $O(1)$, the second-order transitions
described above become first-order. A unified treatment of type-I and type-III transitions on
the global phase diagram, corresponding to weaker and stronger transverse fields respectively,
is an important issue deferred to future studies.

Finally, our concrete calculations using a well-defined model are important for laying a
theoretical foundation for the overall understanding and classification of quantum-critical
heavy-fermion metals. By providing evidence for a line of local critical points on the global
phase diagram, our findings contribute a robust basis for understanding the unusual properties
characteristic of Kondo-destruction quantum criticality observed in various heavy-fermion
compounds. Furthermore, our study suggests a new experiment, namely, to study pressure-induced
quantum criticality in CeCu$_{6-x}$Au$_x$ in the presence of a small but nonzero transverse
magnetic field.

To summarize, we have identified a concrete Kondo lattice model in which the dynamical effects of
tunable quantum fluctuations of the local-moment magnetism can be studied on an equal footing
with those associated with the Kondo coupling.
We have demonstrated, for the first time in a concrete model, a line of fixed points describing
local quantum critical points, where the onset of antiferromagnetic order is concurrent with the
destruction of Kondo entanglement. This result provides a strong basis for the classification
of quantum phase transitions of antiferromangetic heavy-fermion metals in terms of the global phase
diagram, and motivates a new experiment to test our predictions.

{\it Acknowledgements:}
We would like to acknowledge useful discussions with A.\ Cai, A.\ H.\ Nevidomskyy, S.\ Kirchner,
J.\ H.\ Pixley, and J.-X.\ Zhu. This work was supported in part by NSF Grant No.\ DMR-1309531 and
the Robert A.\ Welch Foundation Grant No.\ C-1411 (E.M.N. and Q.S.), and by NSF Grants No.\
DMR-1107814 and No.\ DMR-1508122 (K.I.). The computation was mainly performed, with support from
the U.S.\ Army Research Office Grant No.\ W911NF-14-1-0525  (Q.S.), on the Shared University Grid
at Rice funded by NSF Grant No.\ EIA-0216467, a partnership between Rice University, Sun
Microsystems, and Sigma Solutions, Inc.

\pagebreak
\widetext
\begin{center}
\textbf{\large Quantum criticality and global phase diagram of an Ising-anisotropic Kondo lattice:\\[+2ex] Supplemental Material}
\end{center}

\setcounter{equation}{0}
\setcounter{figure}{0}
\setcounter{table}{0}
\setcounter{page}{1}
\makeatletter
\renewcommand{\theequation}{S\arabic{equation}}
\renewcommand{\thefigure}{S\arabic{figure}}
\renewcommand{\bibnumfmt}[1]{[S#1]}
\renewcommand{\citenumfont}[1]{S#1}

\title{Quantum criticality and global phase diagram of an Ising-anisotropic Kondo lattice:\\[+2ex] Supplemental Material}
\author{Emilian M.\ Nica}

\author{Kevin Ingersent}
\author{Qimiao Si}

\date{\today}
\maketitle

The Supplemental Material is divided into five parts.
We discuss the numerical implementation of the extended dynamical mean-field theory (EDMFT)
of the Kondo lattice model with a transverse field $\Delta$, the large-transverse-field 
regime ($\Delta/T^{(0}_{K} \geq 1$) of our calculations, a previous study of the \emph{standalone}
transverse-field Bose-Fermi Kondo Model (BFKM) that suggested the presence of a line of critical points, 
the extension of the marginality to the self-consistent case, and the dependence of the critical exponents
on the numerical renormalization-group (NRG) discretization parameter $\Lambda$.

\section{I.~Numerical implementation of EDMFT equations}

For given coupling constants of the BFKM [Eq.~(5) in the main text], each iteration of
the numerical solution of the EDMFT self-consistency conditions [Eqs.~(8) and (9)]
involves two steps. Using the current estimates for the Weiss field
$\chi^{-1}_{0}(\omega)$ and the local field $h_{\loc}$, we determine $\chi_{\loc}(\omega)$
and, for ordered solutions, the order parameter $m_{\text{AF}}$. 
The second step involves updating the estimated $\chi^{-1}_{0}(\omega)$ and $h_{\loc}$
through application of the self-consistency conditions. 
The procedure is iterated until the Weiss field satisfies a predefined convergence
criterion.
The first step requires solution of the BFKM. In our calculations, this was achieved
using an extension of the numerical renormalization-group (NRG) method \cite{Wilson_Rev_Mod_Phys_1975} 
to treat impurities coupled both to a fermionic band and to a bosonic bath~\cite{Glossop_Ingersent_PRL_2005}.
The method maps the BFKM to a pair of Wilson chains, one fermionic and the other bosonic, each coupled
at one end to the impurity spin degree of freedom.

Because the NRG calculates the imaginary part of the dynamical spin susceptibility through
broadened delta functions, the $\omega \rightarrow 0$ value of the real part of the spin
susceptibility (calculated via a Hilbert transform)
differs from the result that would arise from
a direct linear-response calculation. Although the mismatch can be partially
corrected~\cite{Glossop_Ingersent_PRL_2007,Zhu_et_al_PRL_2007}, 
it still results in a nominal coexistence
region of the phase diagram [see Figs.~2(a)-(c)
of the main text]. 
It is important to note that the extent of this overlap region for $\Delta>0$
(less than 5\% of the estimated critical coupling) is on the same scale as found
in EDMFT studies for $\Delta=0$~\cite{Glossop_Ingersent_PRL_2007}.
In accordance with this previous EDMFT work, we consider that the coexistence
observed in the present results is fully compatible with a second-order transition at $T=0$.
Moreover, a Landau analysis for the EDMFT equations led to a  continuous transition in
 the $\Delta=0$ case~\cite{Zhu_et_al_PRL_2007}, and this analysis also applies to 
 the nonzero $\Delta$ cases we have presented in the main text.
Consistent with this conclusion, the $T=0$ transition of the standalone BFKM in zero
transverse field was evidenced to be second order
~\cite{Glossop_Ingersent_PRL_2005,Nica_et_al_PRB_2013}.
Furthermore, the critical behavior of the local quantities of the Kondo lattice model,
discussed in the main text, is also consistent with the $\Delta=0$
analytical~\cite{Si_et_al_PRB_2003, Grempel_Si_PRL_2003} and
numerical~\cite{Zhu_et_al_PRL_2003, Glossop_Ingersent_PRL_2007, Zhu_et_al_PRL_2007} studies.
Altogether, these result provide evidence for the second-order nature of the quantum phase transitions. 

All of the results presented in the main text correspond to self-consistent, zero-temperature solutions
of the EDMFT equations, converged to 0.001\% precision in all quantities and for all frequencies
such that $10^{-7} \lesssim \omega/T_K^{(0)} \lesssim 1$, where $T_K^{(0)}=0.476 D$.

\section{II.~Large transverse-field regime within the EDMFT}

In the limit of $J_K \rightarrow 0$ ($T_{K}^{(0)} \rightarrow 0$), the Kondo lattice model (KLM)
[Eq.~(1)] reduces to the transverse-field Ising model (TFIM)~\cite{TFIM}
which exhibits second-order transitions between Ising long-range ordered and spin-gapped
paramagnetic phases. Inside the latter phase with small but nonzero $J_K$, we expect
the emergence of an analogous $P_S$ phase~\cite{Yamamoto07}. In practice, this regime is
difficult to study within the EDMFT. To illustrate, we consider the $\Delta$-induced
transition at $T=0$, which---in the absence of any coupling to fermions---has a 
dynamic exponent $z=1$. For spatial dimensions $d=2$, the dynamics raise the effective
dimensionality to $d_{\text{eff}}=d+z=3$, which is less than the upper critical dimension of $4$.
As a result, a spatial anomalous dimension must emerge.
In this case, the EDMFT equations are expected to yield a first-order phase transition, as
demonstrated by the transition at nonzero temperature $T_N$ in the EDMFT treatment of the KLM
(where $d_{\text{eff}}=2$, also below the upper critical dimension)~\cite{Zhu_et_al_PRL_2003}.
For the full KLM ($J_K, \Delta \neq 0$), we expect this spurious effect to come into play for
$\Delta/T_K = \text{O}(1)$. Indeed, for the case $\Delta^{(0)}=0.63$ (not shown) we obtain a large
coexistence region between completely screened and finite magnetic order-parameter solutions
with a width approaching 100\% of the estimated $\delta_{c,\text{AF}}$. We stress that, for
the range of $\Delta/T^{(0)}_{K}$ values discussed in the main text,  we have provided
ample evidence for a line of \emph{continuous} zero-temperature phase transitions.

\section{III.~Marginal behavior and line of unstable fixed points in the BFKM without self-consistency}

In anticipation of the self-consistent solution for the BFKM with a transverse field,
an NRG study was previously undertaken of a standalone (non-self-consistent) BFKM
with a sub-Ohmic bosonic spectral function $B(\omega)$~\cite{Nica_et_al_PRB_2013}.
The results for this independent model strongly suggest the existence of a line of critical
points generated by the transverse field. Equivalently, $\Delta$ can be considered to be
a marginally relevant perturbation about the $\Delta=0$ fixed point. Since the self-consistent
results must belong to a subset of the standalone solutions (see discussion on scaling analysis
below), our starting hypothesis in the self-consistent case was that a marginal scaling variable
also emerges in the EDMFT approach. The self-consistent results presented in the main text are in
agreement with this scenario.

We give a brief overview of the previously obtained~\cite{Nica_et_al_PRB_2013} NRG results for
the standalone BFKM coupled to a transverse field with the bosonic spectral function
$B(\omega)$ chosen to have the power-law form
\begin{equation}
\label{Eq:B_sngl_imprt}
B = g_0 \left( \frac{\omega}{\omega_0} \right)^s \Theta(\omega) \, \Theta(\omega_0 - \omega),
\end{equation}
where $g_0$ is an effective coupling to the bosons, $\omega_0$ is a high-energy cutoff scale, and
$s=0.8$. (Similar results are obtained for other sub-Ohmic exponents $s<1$).
In an EDMFT setting, such a spectral function would be characteristic of a self-consistent
Weiss field approaching a critical dynamical form. In the standalone model, a line of critical
points is found to separate screened and localized moment phases in the presence of a transverse
field. Since self-consistency is not imposed, the critical behavior is much more accessible and
solutions with $\left(g-g_c\right)/g_c$ of O$(10^{-6})$ are readily obtained. In this proximity
to the critical manifold, and in an intermediate scaling regime, the NRG spectrum for Wilson
discretization parameter $\Lambda=9$ shows clear invariance over roughly 7-10 NRG iterations,
corresponding to a rescaling by a factor of $\Lambda^{1/2}$ per iteration or roughly 4-5
decades overall. Over this energy window, the NRG spectrum can be
presumed to provide an accurate approximation to the spectrum of the critical fixed point
Hamiltonian~\cite{Wilson_Rev_Mod_Phys_1975}. 
Within the NRG, two fixed points are identical if, for the same value of $\Lambda$, they
share the same spectrum as well as equivalent matrix elements of every local operator
between many-body eigenstates. Thus, comparison of NRG spectra provides a
powerful means of studying the fixed-point structure of a problem.

\begin{figure}[t!]
\subfloat[]{\includegraphics[width=0.42\columnwidth]{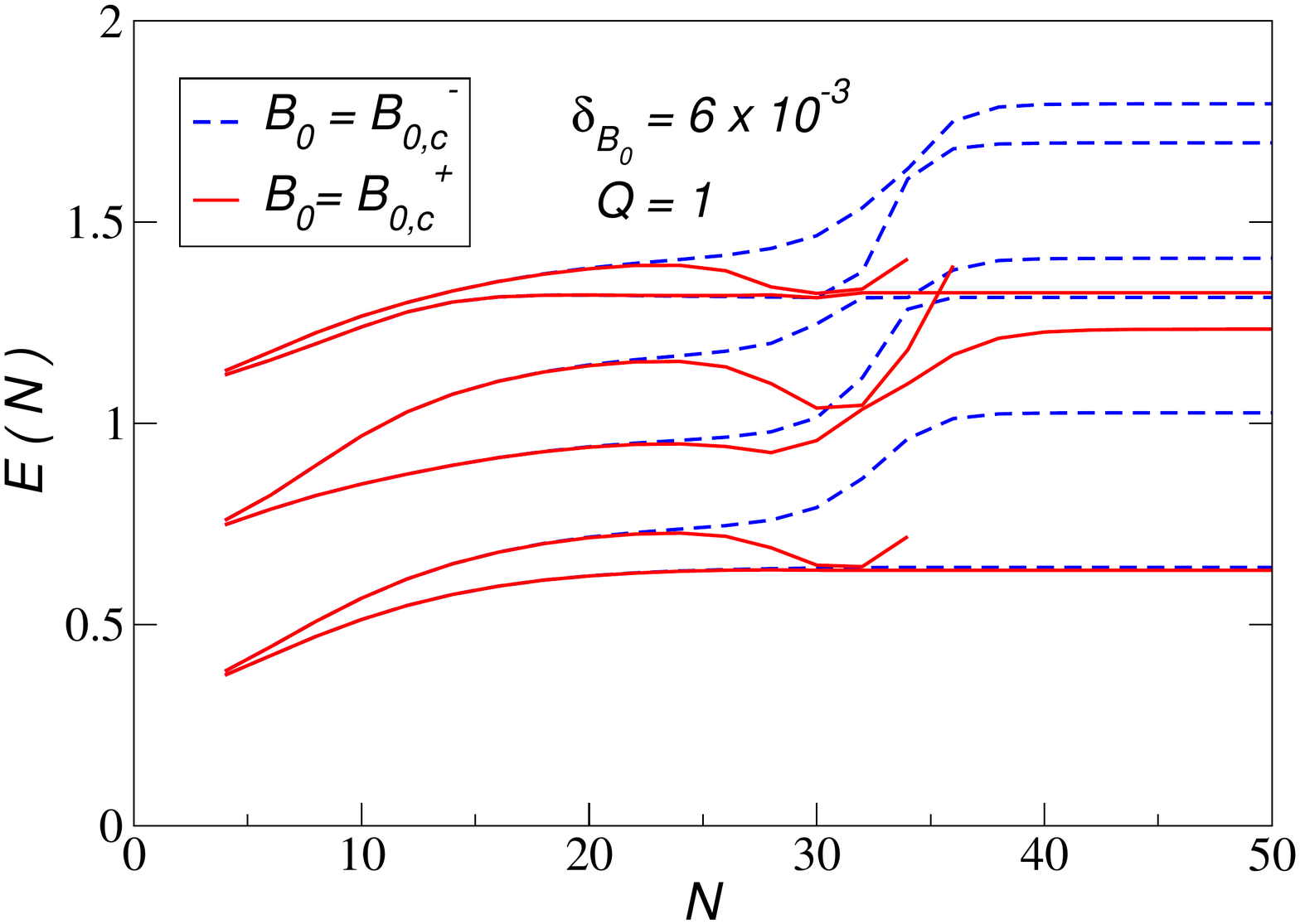}} \\[-4ex]
\subfloat[] {\includegraphics[width=0.42\columnwidth]{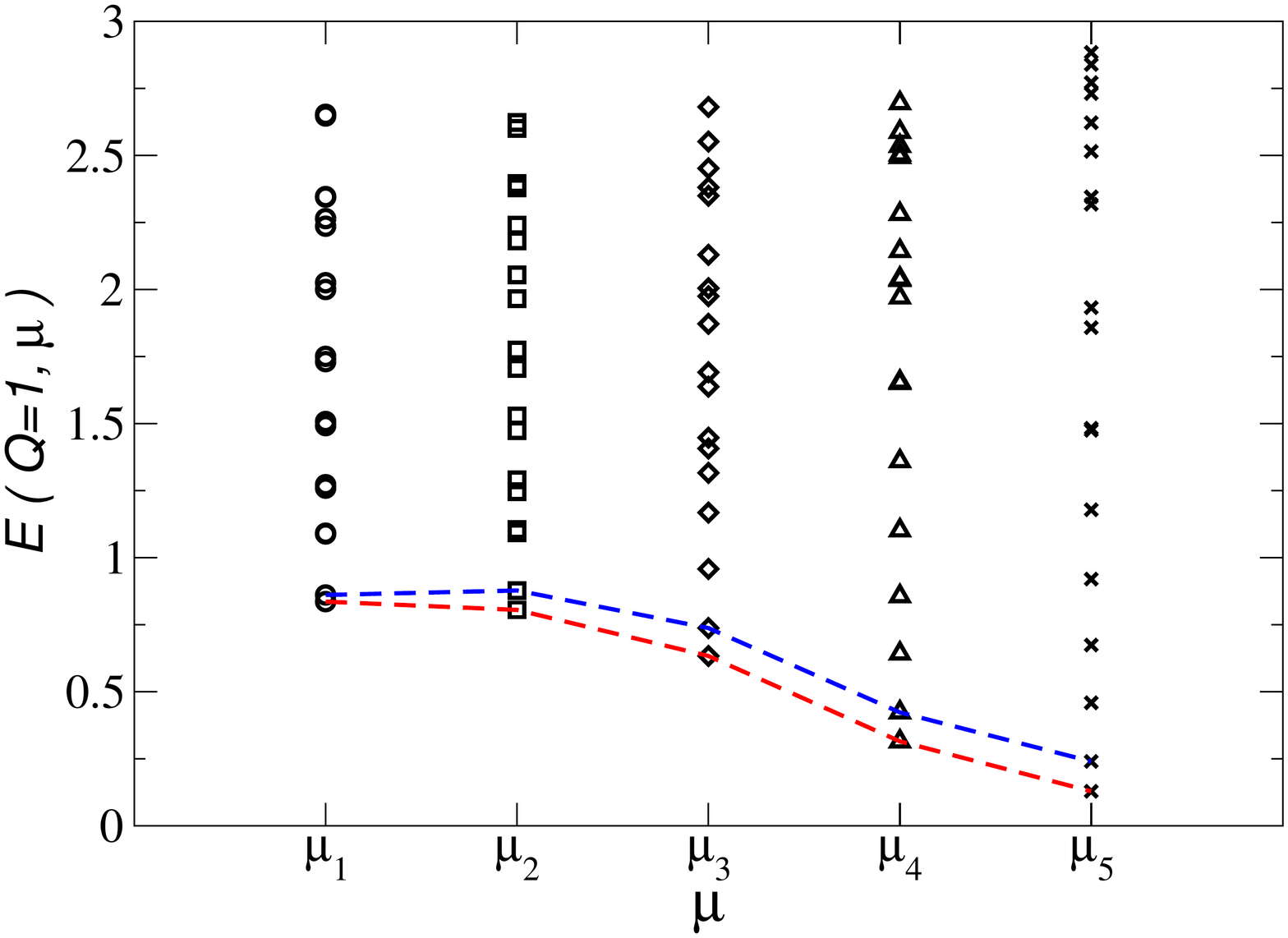}} \\[-4ex]
\subfloat[]{\includegraphics[width=0.42\columnwidth, height=0.5\columnwidth]{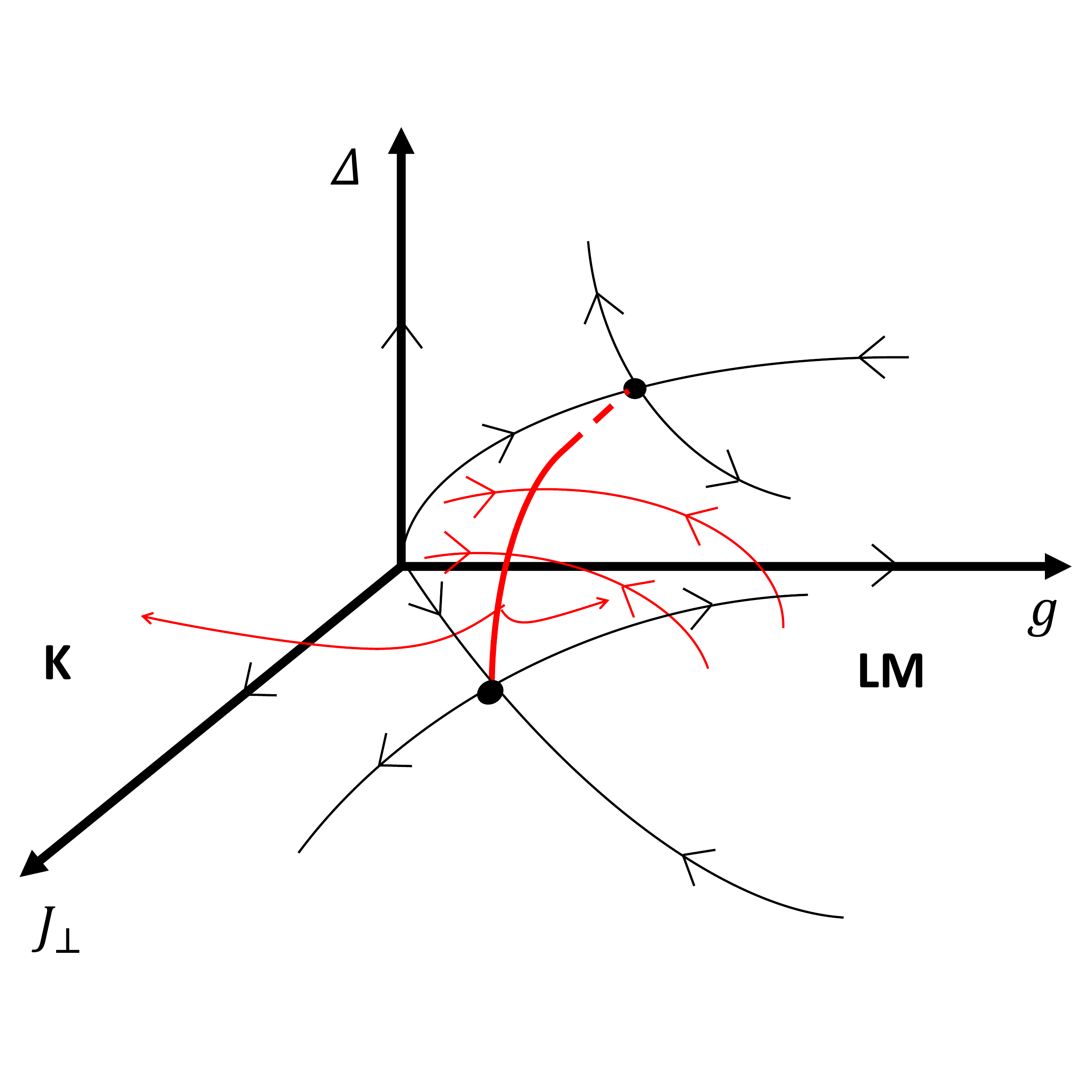}}
\caption{(a) Eigenenergy $E$ vs NRG iteration number $N$ for the six lowest $Q = 1$ states of the
BFKM for the same transverse field $\Delta \neq 0$ and two near-critical bosonic couplings satisfying
$\delta_{B_0} = |g-g_c|/g_{c} = 6 \times 10^{3}$: one (labeled $B_{0,c}^-$) on the screened side and
the other ($B_{0,c}^+$) on the localized side. For iterations $N>24$, the spectra flow away from the
critical spectrum toward the localized moment (solid lines) and screened moment (dashed) stable fixed
points. (b) Lowest $Q = 1$ eigenvalues closest to the critical point ($N \simeq 24)$ for values of
$\Delta$ that increase from $\mu_1$ to $\mu_5$ along the horizontal axis. (See
Ref.~\onlinecite{Nica_et_al_PRB_2013} for details). The dashed lines track the evolution of the splitting
between the two lowest states. (c) Schematic renormalization-group flow diagram for a generic BFKM
in a space spanned by the running values of the spin-flip Kondo coupling $J_{\perp}$, the bosonic
coupling $g$, and the transverse magnetic field $\Delta$. $K$ (Kondo) and $LM$ (local-moment) represent
the asymptotic regimes where the local moment is either screened or has a finite expectation value,
respectively. The arrows indicate the relevant (outbound) and irrelevant (inbound) directions
near each fixed point. The thick red line represents a line of unstable fixed points which
emerges from the $\Delta=0$ fixed point as the transverse field is turned on. The red arrows show
flows within and away from the critical surface that separates the screened and localized
phases. As the shape of the critical surface is expected to be analytical with with respect to
$\Delta$, for small transverse fields the flows must closely follow the in-plane $\Delta=0$
flows. All panels are reproduced from Ref.~\onlinecite{Nica_et_al_PRB_2013}.}
\label{Fig:Critical_spectrum}
\end{figure}

In a transverse field, the critical spectra discussed above exhibit clear Zeeman-like
splitting~\cite{Nica_et_al_PRB_2013}. This is clearest for pairs of states with
an odd value of the total conserved charge (measured from half-filling)
\begin{equation}
\label{Eq:Chi_ln_scaling}
Q=\sum_{n=0}^{N} \left( \sum_{\sigma} f^{\dagger}_{n, \sigma} f_{n, \sigma} -1 \right),
\end{equation}
where $f_{n\sigma}$ annihilates an electron of spin index $\sigma=\pm\frac{1}{2}$
at site $n$ of the Wilson chain, and $N=0$, $1$, $2$, $\ldots$ is the NRG iteration number.
For these odd-$Q$ pairs, the splittings increase with $\Delta$. 
The plateaus in the $Q=1$ spectrum vs $N$ around $N=24$ are illustrated
in Fig.~\ref{Fig:Critical_spectrum}(a), while the size of the splittings for the
two lowest $Q=1$ levels are plotted as functions of $\Delta$ in
Fig.~\ref{Fig:Critical_spectrum}(b).

\noindent 

The same effect is harder to ascertain for even values of $Q$. However, there are indications that
for $Q=0$, the lowest part of the spectrum is independent of $\Delta$,
while higher states, possibly mixing with $Q=0$ particle-hole excitations of the fermions, do vary
with transverse field. The continuous variation of some of these states with increasing $\Delta$
naturally suggests the presence of a marginal scaling variable which generates a line of unstable fixed
points. The different behaviors of the even-$Q$ and odd-$Q$ spectra also suggest that the marginal
operator is mainly associated with fermionic excitations. 

There are indications of the existence of such a marginal operator from analytic weak-coupling 
Coulomb-gas studies of similar models~\cite{Vladar_et_al_PRL_1986, Moustakas_Fisher_PRB_1995} 
such as the two-state system (spin-boson model) in the presence of Kondo-like exchange couplings. 
In particular, Ref.~\onlinecite{Moustakas_Fisher_PRB_1995} showed that the equivalent of the
transverse field, together with the spin-dependent longitudinal Kondo interactions generate
a marginal coupling under the renormalization group (RG) that locally hybridizes spin up/down
conduction electrons. This effect introduces spontaneous ``spin-flip''
for the conduction electrons or equivalently spin-density fluctuations which reduce the screening
(``orthogonality'' effect) of the pure Kondo spin-flip scattering. Unfortunately, 
these studies do not consider a coupling to a bosonic bath that can lead to corrections to scaling. 
The possible extension of these works to the full BFKM remains an important future direction. 
Here, we note that if such an operator were to remain marginal at the $\Delta=0$ critical point, 
it could pinpoint a way to understand both the standalone and self-consistent BFKM numerical results.
Physically, we expect that marginal couplings between the conduction electrons 
are likely to cause a gradual decrease in the electrons' ability to screen the impurity 
and to introduce additional local critical modes.  

The line of unstable fixed points discussed above is illustrated by the red line in the RG flow diagram
shown in Fig.~\ref{Fig:Critical_spectrum}(c).
The appearance of a continuous line extending from the $\Delta=0$ unstable fixed point implies the
existence of an operator associated with the transverse field that is marginal (at least over
a range of small $\Delta$).

\section{IV.~Marginal behavior in the self-consistent solutions}

In the previous section we discussed the emergence of a line of unstable fixed points for the
\emph{standalone} BFKM with finite values of the bare transverse field. Here, based on the general
correspondence between the standalone and self-consistent versions of the BFKM, we argue that
a similar line of fixed points appears in the solutions of the EDMFT equations, \textit{i.e.},
that a finite transverse-field introduces an additional marginal operator in the lattice case as well.

As shown in Refs.~\onlinecite{Si_et_al_PRB_2003, Grempel_Si_PRL_2003}, for the self-consistent BFKM
with $\Delta=0$, the leading logarithmic scaling form of the local critical susceptibility is
determined by the self-consistency condition [Eq.~(9)]. Up to irrelevant terms, the correspondence
between the leading low-frequency form of the spectral function $B(\omega)$ [Eq.~\eqref{Eq:B_sngl_imprt}]
and the Weiss field $\chi_{0}^{-1}(\omega)$ implies that self-consistent critical solutions form a subset
of the \emph{standalone} model.

On general grounds, we expect that self-consistent critical solutions are also a subset of solutions
of standalone models both for zero and nonzero transverse fields. The results presented in the main text indicate that for
certain critical combinations of the couplings $J_K, I$, and $\Delta$, the system exhibits the dynamical logarithmic form of the zero 
transverse-field solution. Two possibilities then arise: either the set of self-consistent critical solutions is mapped onto a single effective
$\Delta=0$ unstable fixed point or, for each value of the transverse-field, the RG flows are determined by different fixed points. In either case, the leading critical scaling must be compatible with that of the $\Delta=0$ solution, with an expected renormalization of nonuniversal scales. Equivalently, $\Delta$ is either irrelevant about the zero-transverse field unstable fixed point or it introduces a marginal scaling variable. In view of the results of the standalone BFKM with $\Delta>0$, we believe that the latter scenario is more likely.
The superscaling shown in Figs.~4(a) and 4(b) is consistent with the presence of a marginal operator. 

In practice, the distinction between irrelevant and marginal cases is difficult to establish unambiguously due to the heavy computational demands in the vicinity of the critical point. The difference between the two possibilities is important from a physical point of view. In the irrelevant case, the fluctuations of the local moments, parameterized by $\Delta$, cannot tune a relevant scaling variable
that allows any transverse field (however large) to
modify the critical low-energy theory. In the marginal case, as discussed in the previous section, in the progressive splitting of the spectrum with increasing $|\Delta|$ suggests a gradual disentanglement of the critical modes associated with Kondo destruction, and an evolution to a fixed point increasingly dominated by the fluctuations of the local moment. \emph{A priori}, this makes possible the $AF_S \leftrightarrow P_S, P_L \leftrightarrow P_S$ transitions that are expected for large transverse fields.

For completeness, we also discuss the possibility that $\Delta$ introduces an additional relevant operator. Close to the self-consistent
$\Delta=0$ fixed point, the modified RG flows become unstable in a new direction, eventually heading to a different fixed point. If this point is stable, the second-order transition must become a cross-over. Alternately, a second \emph{unstable} fixed point would most likely generate a new dynamical critical form. 
The fact that the local susceptibility appears to maintain its $\Delta=0$ logarithmic form, the static superscaling shown in Fig.~4(b), and the natural expectation of marginal behavior discussed above, all make this scenario unlikely.

\section{V.~Dependence of critical exponent $\alpha$ on NRG discretization parameter $\Lambda$}

Both in its original form~\cite{Wilson_Rev_Mod_Phys_1975} and its Bose-Fermi (two-chain) extension~\cite{Glossop_Ingersent_PRL_2005},
the NRG involves a discretization parameter $\Lambda>1$.
Its essential role is to provide an artificial separation of energy scales along the Wilson chain representation of the conduction band
and of the bosonic bath. The values of certain properties calculated using the NRG results depend on $\Lambda$, and the physical values must be obtained via extrapolation to the continuum limit $\Lambda \rightarrow 1$. The exponent $\alpha$ entering Eq.~(10) is one such property.
An NRG treatment of a pure-bosonic mapping of the self-consistent BFKM in zero transverse field discovered a logarithmic
dependence of $\alpha$ on $\Lambda$~\cite{Glossop_et_al_unpub_2013}.
The good fit in that case (based on multiple $\Lambda$ values spanning half a decade) motivated the analogous
procedure used in the present work, where the extrapolation to $\Lambda=1$ was based solely on data points
for $\Lambda=3$ and $\Lambda=9$.

\end{document}